\documentclass[manuscript]{aastex}

\begin{document}

\shorttitle{V2552 Oph}
\shortauthors{Hesselbach et al.}

\title{The Newly Active R Coronae Borealis Star, V2552 Oph}

\author{E. Hesselbach\altaffilmark{1,2,3}, Geoffrey C. 
Clayton\altaffilmark{1,4},and
Paul S. Smith\altaffilmark{5}}

\altaffiltext{1}{Maria Mitchell Observatory, 3 Vestal St. Nantucket, MA 
02554}
\altaffiltext{2}{Dept. of Physics, Alfred University, Alfred, NY 14802; 
hesselen@alfredalum.com}
\altaffiltext{3}{Now at Ritter Observatory, University of Toledo,
2801 W. Bancroft, Toledo, OH 43606}
\altaffiltext{4}{Department of Physics and Astronomy, Louisiana State 
University, Baton Rouge, LA 70803; gclayton@fenway.phys.lsu.edu}
\altaffiltext{5}{Steward Observatory, University of Arizona, Tucson, AZ 85721;
psmith@as.arizona.edu}

\begin{abstract}
In 2001, V2552 Oph (CD -22 12017, HadV98) quickly faded by several
magnitudes in a manner typical of the R Coronae Borealis (RCB) stars. 
Photometry of V2552 Oph  
obtained over 70 years previous to 2001 shows no
indication of variability.
Optical spectra
of this star subsequently confirmed that V2552 Oph is a member of the 
hydrogen deficient, carbon-rich 
RCB class of variables. It resembles the warm (T$_{eff}\sim$ 7000 K) RCB stars
such as R Coronae Borealis itself. Other RCB stars, such as XX Cam and Y Mus,
have experienced similar periods of inactivity, going decades without 
significant dust 
formation.
Further observations of V2552 Oph will be of great interest since there is an 
opportunity to monitor an RCB star that may be moving from prolonged 
inactivity into an active 
phase of
dust production.
\vspace*{0.2in}
\end{abstract}


\section{Introduction}

	R Coronae Borealis (RCB) stars are rare, hydrogen-deficient, 
carbon rich 
supergiants (Clayton 1996).  At irregular intervals, they exhibit dramatic 
declines of 2 to 8 magnitudes.  These declines are caused by dust formation 
along the line of sight.  The rarity of RCB stars may stem from the fact that 
they are in an extremely rapid phase of the evolution toward white dwarfs.  
Therefore, understanding RCB stars is a key test for any theory attempting to 
explain hydrogen deficiency in post-Asymptotic Giant Branch (AGB) stars.  
Only about 35 RCB stars are known in the Galaxy.  Because of their rarity, 
the identification of new RCB stars is important to the study of this exotic 
class of star. 
RCB stars are true irregular variables 
(Clayton, Whitney, \& Mattei 1993 and references therein).  
They spend a majority of their 
time at maximum 
light with a characteristic time between declines of about 1100 days 
(Feast 1986). However, there is a wide variation in decline activity from star
to star (Jurcsik 1996). Also, individual stars often vary from extremely 
active to extremely inactive on the timescale of decades (e.g., 
Mattei, Waagen, \& Foster
1991,1993).

On the basis
of a more complete census of RCB stars in the LMC than is possible in the 
Galaxy, Alcock et al. (2001) suggested that many RCB stars
lie undiscovered in the 
Galaxy. They also suggest that cool (T$_{eff}\sim$ 5000 K) RCB stars are far 
more 
numerous but intrinsically fainter than the warm (T$_{eff}\sim$ 7000 K) stars. So these fainter, redder stars may have been selectively missed in 
previous surveys. This suggestion has been supported in that RCB stars recently
discovered in the Galaxy have tended to be cool (e.g., Clayton et al. 2002; 
Benson et al. 1994).

In 2001, the star CD -22 12017, not previously known to be a variable, went
into a deep RCB star-like decline (Haseda \& Kato 2001). This new variable
star was temporarily known as HadV98 and is now officially designated as
V2552 Oph (Kazarovets et al. 2003).
This paper presents photometric and spectroscopic data suggesting that 
V2552 Oph is indeed a member of the RCB class of 
variable stars. 
 V2552 Oph is listed in the Hubble Space Telescope Guide Star Catalog (GSC 2.2
S223121279620)
with coordinates, $\alpha$(2000.0) = 17$^h$ 23$^m$ 14\fs556, 
$\delta$(2000.0) = 
-22\arcdeg~52\arcmin~06\farcs23.

\section{Spectroscopy}

We obtained a low resolution spectrum of V2552 Oph on 
2002 June 10 using the Steward Observatory 90-inch~telescope with its 
Boller \& Chivens spectrograph, a CCD detector, 4\farcs5~slit 
and a 300 line $mm^{-1}$ 
grating. 
A spectral range from 4300 to 
8200 \AA\ was recorded at a spectral resolution of 
about 10 \AA. 
The spectrum is shown in Figure 1 along with a spectrum of the
LMC RCB star, W Men for comparison (Alcock et al. 2001). The spectra are
not photometric so the slope differences should not be given credence.

The spectra of the two stars are very similar. 
They are very typical warm (T$_{eff}\sim$ 7000 K) RCB star spectra having 
the
following features: weak or absent hydrogen lines and CH bands, strong
carbon lines and fairly weak molecular bands (CN, C$_2$) (Clayton 1996). 
These features are marked in Figure 1.
The identification of V2552 Oph as an RCB star has been confirmed 
recently by comparing
a high resolution spectrum of V2552 Oph to the spectra of other 
RCB stars (Rao \& Lambert 2003). 
They find that the spectrum is very 
similar to that of R CrB itself, and 
that the abundances of V2552 Oph most closely resemble FH Sct and 
Y Mus in that Sr, Y and Zr are enhanced.

\section{The Lightcurve}

The recent lightcurve of V2552 Oph from 1997 to 2003 is shown in Kato \& 
Haseda (2003).  The star was apparently constant in brightness at 
V$\sim$10.7 mag for almost 4
years from JD 2450654 to 2452022. Twenty days later, V2552 Oph was discovered 
to be 
in a deep decline 
at V $\sim$12 mag (Kato \& Haseda 2003). Before 1997, the brightness
of V2552 Oph was not monitored regularly. It was observed by Hipparcos in 
1991. The Tycho-2 catalogue gives V=10.9 $\pm$ 0.2 mag and B-V =1.0 $\pm$ 0.2 
mag. These
magnitudes have been converted to the Johnson system (ESA 1997) 
and are consistent with the maximum light 
brightness of V2552 Oph given by Kato \& Haseda (2003).
Rao \& Lambert (2003) estimate the reddening toward V2552 Oph to be 
E(B-V)$\sim$0.4 mag based on the strength of a diffuse interstellar band. The
measured B-V of V2552 Oph is $\sim$1.0 mag. Therefore, its intrinsic B-V would be
0.6 mag in good agreement with the intrinsic colors of other warm RCB stars 
such as R CrB and RY Sgr. 

V2552 Oph appears in the first volume
of the Cordoba Durchmusterung (CD) catalog with a visual brightness of 9.7 mag
(Thome 1892).  
The transformation from $m_{vis}$ to $V$ has been determined to be,
$m_{vis} - V$  = 0.21$\times$($B-V$) (Stanton 1999). So for V2552 Oph, the
CD catalog implies that V2552 Oph had a brightness of V$\sim$9.5 mag sometime
before 1892.
The star does not appear in any other stellar catalog including
the HD, CP and SAO catalogs.  The fact that V2552 Oph does not appear in 
these catalogs makes it likely that its magnitude and color were more 
similar to their present day values than the uncertain V$\sim$9.5 mag quoted 
in the CD 
catalog.  
V2552 Oph was not noticed to be variable at any time
before 2001.

The Harvard College Observatory (HCO) 
Photographic Plate collection was searched 
for fields 
containing V2552 Oph.  The majority of the HCO collection consists of 
blue plates taken with a 
variety of refractors
having focal scales of 60-600 $\arcsec~mm^{-1}$.  Eye estimates of the 
photographic magnitude (m$_{pg}$) of V2552 Oph were carried out on 
80 plates covering 1931-1951 and 1970-1989.  The temporal coverage 
of the HCO plates is plotted in
Figure 2. The 
brightness of V2552 Oph was visually compared to that of two neighboring 
stars of similar magnitude, GSC2.2 S223121279668   
and S223121280800. 
No significant 
brightness variations were found. On all plates 
studied V2552 Oph was observed to be constant at or near maximum light, 
at around m$_{pg}$=12.2 mag.  The 
uncertainty in the $m_{pg}$ measurements is of order $\pm$0.4~mag.
The 
transformation 
from $m_{pg}$ to 
Johnson $V$ 
($V-m_{pg}$ = 0.17 - 1.09$\times$($B-V$)) is $V-m_{pg}$ = -0.9 
mag (Arp 1961; Pierce \& Jacoby 1995). So the brightness of V2552 Oph
on the HCO plates obtained between 1931 and 1989 is
V$\sim$11.3 mag, which is consistent
with its maximum light brightness at the present epoch. 
V2552 Oph is also present on various survey plates (e.g., USNO, GSC, SERC) 
obtained on the following dates:
1950.529, 1980.485, 1986.940, 1987.579, 1987.634, 1987.650, 1992.555, 1992.57, 1997.244,
and 1999.490.  
On all of these plates, V2552 Oph appears to be at maximum light.
The lightcurve of V2552 Oph, while fragmentary, shows no sign of a decline
in the 70 years before 2001. It is certainly possible for 
a small decline of a few tenths of a magnitude to have been missed. 
But a significant decline of several magnitudes will typically last a
year or more.  A good example is the decline that V2552 Oph experienced in 2001-2002 (Kato \& Haseda 2003).
It is unlikely that such a decline was missed during the years for which
at least one HCO plate was measured.

\section{Infrared Photometry}

V2552 Oph has been observed twice in the near-IR by 2MASS and DENIS. In 1999
(2451305.8021),
2MASS found V2552 Oph to be, J=8.69$\pm$0.02 mag,
H=8.39$\pm$0.04 mag, and
K=8.16$\pm$0.02 mag (Cutri et al. 2003).
In 2000 (2451827.983647) DENIS found J=8.68$\pm$0.06 mag and
K=8.14$\pm$0.06 mag. The data agree well within the errors. 
V2552 Oph, when corrected for reddening, lies very close to Y Mus in 
the (H-K)$_O$ Vs (J-K)$_O$ plot (Feast 1997). Both
stars lie in the region of the diagram populated by the hydrogen-deficent 
carbon stars which 
make little or no dust. 
In over a hundred years, Y Mus has been observed to decline only twice, in 
1897 and 1953 
(Feast 1997).
Most RCB stars show an near-IR excess even if they have not
had a decline for several years, indicating that dust is still being formed
near the star just not along the line of sight to the star (Feast 1997). 
However, for
Y Mus and V2552 Oph, the data indicate that recent dust formation has been 
small or non-existent. 
In the 2MASS and DENIS photometry, V2552 Oph exhibited no excess 
emission because the data were 
obtained in 1999-2000 just before the start of
dust production resulted in the decline in 2001. 
This indicates that like Y Mus, V2552 Oph had not been making 
dust recently. 

V2552 Oph is not in the IRAS Point Source Catalog or Faint Source Catalog 
(FSC).  
However,
it does appear in the FSC Rejects list which includes objects with S/N greater
than 3 that do not meet other criteria for inclusion in the FSC. 
V2552 Oph had measured fluxes of 0.27$\pm$0.08 Jy, 0.26$\pm$0.06 Jy, 
1.46$\pm$0.34 Jy, and 11.48$\pm$2.64 Jy in the 12, 25, 60 and 100 
\micron~bands, respectively. In particular, 
the IRAS fluxes for V2552 Oph increase to higher wavelengths. All of the
RCB stars observed with IRAS show the opposite behavior. They are brightest at 
12\micron~(Walker 1985). In particular, Y Mus, shows a 12\micron~flux of
1.02$\pm$0.09 Jy. The estimated distances of Y Mus and V2552 Oph are very
similar assuming they  both have M$_V$= -5 mag (Alcock et al. 2001). Y Mus is
at 7.2 kpc and V2552 Oph is at 7.9 kpc assuming interstellar dust extinction
with R$_V$=3.1 (Cardelli, Clayton, \& Mathis 1989; Lawson et al. 1990). 
If both stars had identical
dust clouds surrounding them, then Y Mus would appear 
about 20\% brighter.  But 
even if the suspect V2552 Oph IRAS flux is correct, Y Mus is 
about 4 times as bright at 12\micron~indicating more dust is present 
around that star. So V2552 Oph may have been inactive for a longer period 
of time than Y Mus.

Spectroscopically, V2552 Oph is a typical warm RCB star. But it is one of the
most inactive of the RCB stars over the 
last 70 years putting it in a class with Y Mus. 
Now that V2552 Oph may have entered an
active phase, it will be interesting to monitor its new dust affects the
near- and far-IR emission from the star especially now that SIRTF has been
launched.

\acknowledgements
We would like to thank Vladimir Strelnitski, Regina Jorgenson, and the 
Maria Mitchell Observatory for their hospitality.
This project was supported by the NSF/REU grant AST-0097694 and 
the Nantucket Maria Mitchell Association.
PSS acknowledges support from NASA and JPL through SIRTF Science Working
Group contract 959969.
We would also like to thank Ed Anderson, Alison
Doane, and Albert
Jones.
This publication makes use of data products from the Two Micron All Sky 
Survey, which is a joint project of the University of Massachusetts and 
the Infrared Processing and Analysis Center/California Institute of 
Technology, funded by the National Aeronautics and Space Administration and 
the National Science Foundation.

\begin{figure*}
\begin{center}
\epsscale{1.0}
\plotone{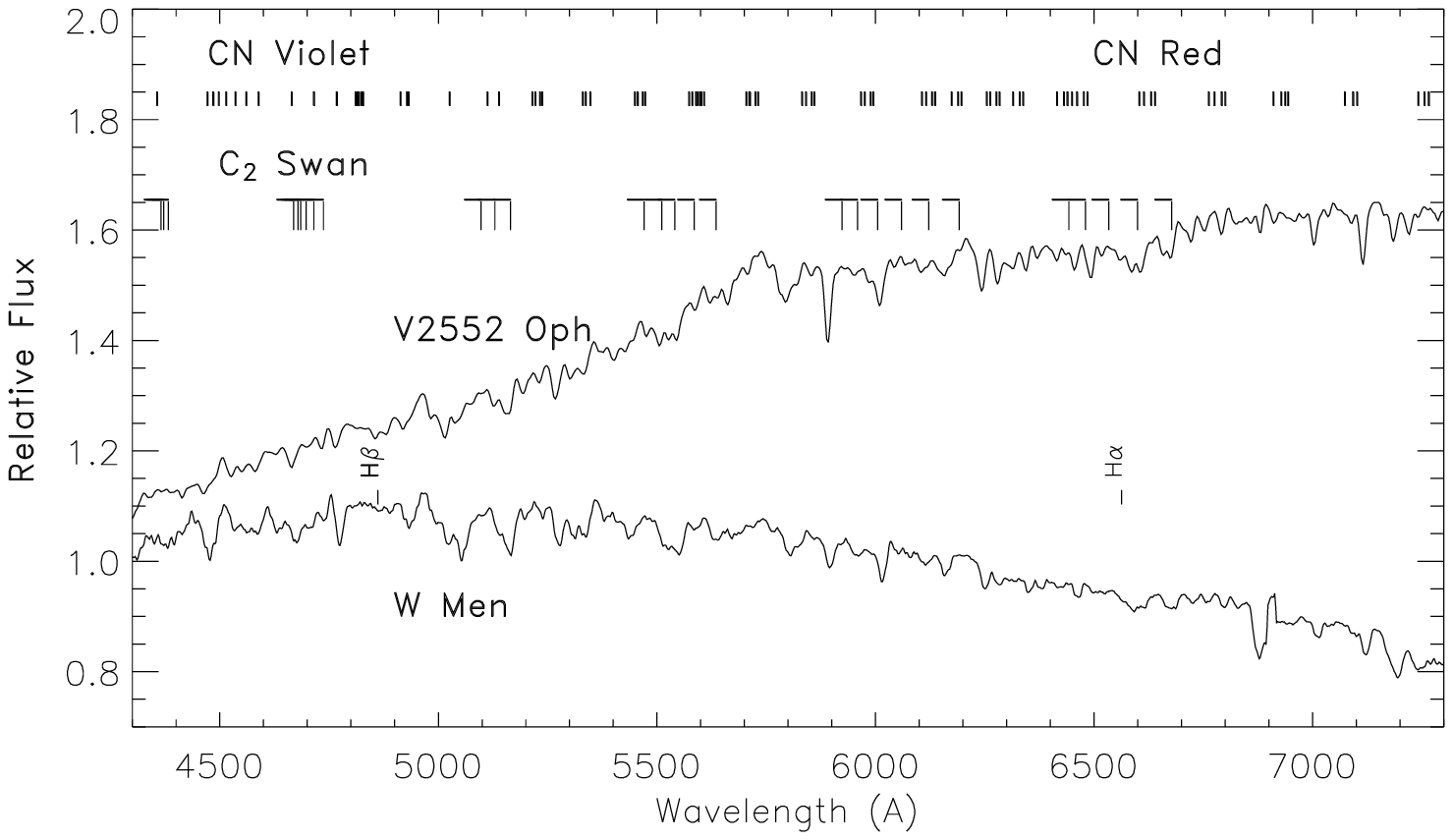}
\end{center}
{\footnotesize Figure 1 -- The spectrum of V2552 Oph taken at maximum light.
Also plotted is the maximum light spectrum of the RCB star, W Men. The flux
levels of both stars have been scaled for easy comparison. The flux
at 6000 \AA\ is $\sim$5.9 x 10$^{-14}$ erg~cm$^{-2}~s^{-1}~\AA^{-1}$ 
for V2552 Oph and 
$\sim$5.8 x 10$^{-15}$
erg~cm$^{-2}~s^{-1}~\AA^{-1}$ for W Men.}
\vspace{0.1in}
\end{figure*}

\begin{figure*}
\begin{center}
\epsscale{1.0}
\plotone{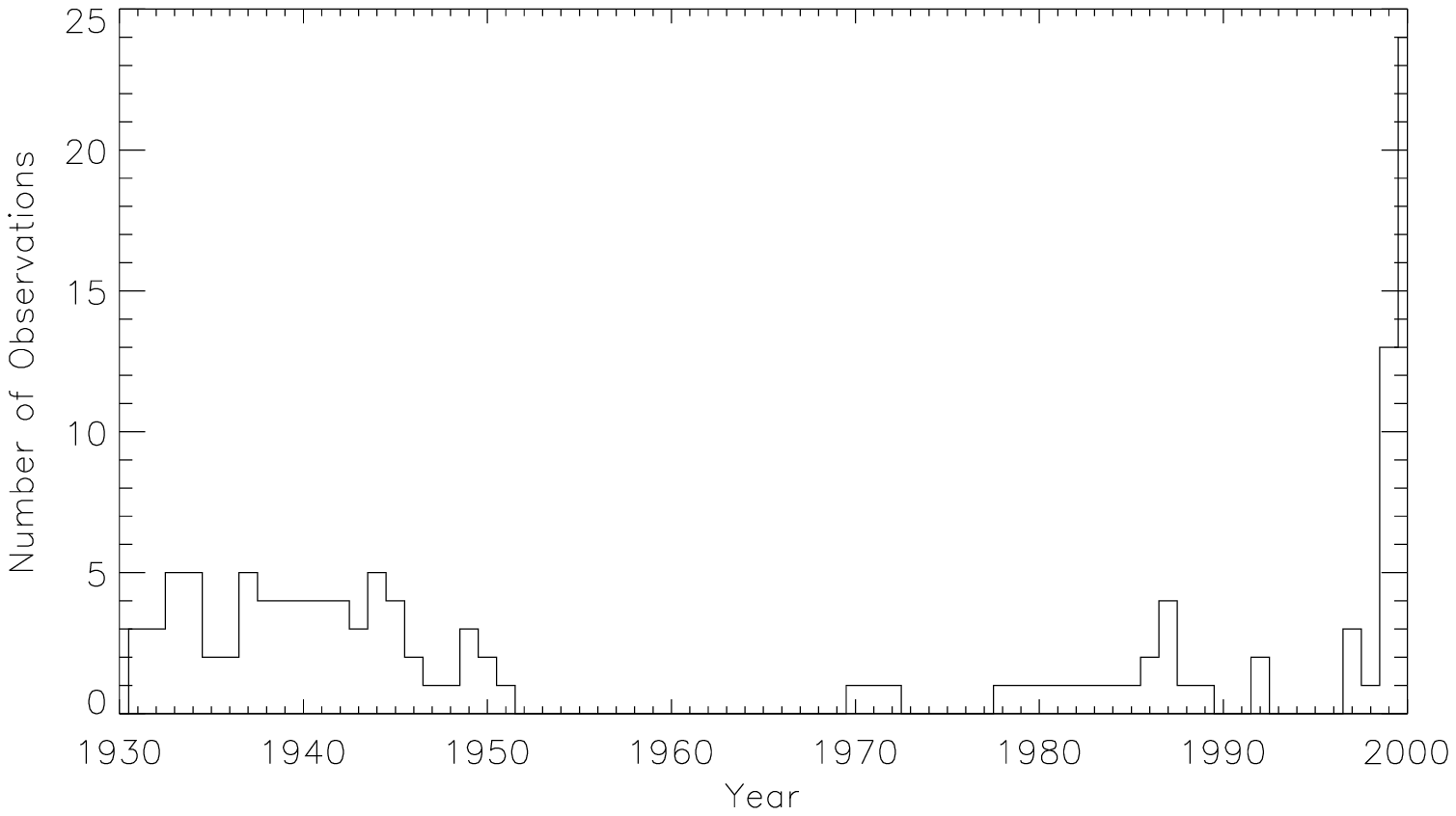}
\end{center}
{\footnotesize Figure 2 -- The number of photometric observations per year of
V2552 Oph obtained from archival photographic plates.
See text.}
\vspace{0.1in}
\end{figure*}

\end{document}